\documentclass[11pt,superscriptaddress]{article}
\usepackage{jheppub}
\usepackage{epsfig}
\usepackage{amsmath,amssymb}
\usepackage{amsfonts}
\usepackage{mathrsfs}
\usepackage{relsize}
\RequirePackage{xspace}
\usepackage{graphicx}
\usepackage{slashed}
\usepackage{bm}

\usepackage{multirow}
\usepackage{graphicx}
\usepackage{bm}
\usepackage[T1]{fontenc}
\usepackage{fourier}

\title{\boldmath $B^*_{s,d} \to \mu^+ \mu^-$~ and its impact
 on $B_{s,d} \to \mu^+ \mu^-$}

\author[a,b]{Guang-Zhi Xu }
\emailAdd{ still200@gmail.com}
\author[b]{, Yue Qiu }
\emailAdd{ qyxy16@gmail.com}
\author[b, c]{, Cheng-Ping Shen }
\emailAdd{shencp@buaa.edu.cn}
\author[b, c]{, Yu-Jie Zhang}
\emailAdd{nophy0@gmail.com}

\affiliation[a]{Department of Physics, Liaoning University, Shenyang 110036
, China}

\affiliation[b]{School of Physics,
  Beihang University, Beijing 100191, China}

\affiliation[c]{CAS Center for Excellence in Particle Physics, Beijing 100049, China}

\abstract{  The decay of
$B^*_{s,d} \to \mu^+ \mu^-$~ and its impact on $B_{s,d} \to \mu^+ \mu^-$ is studied here. The $ \mu^+ \mu^-$ decay widths of vector mesons $B^*_{s,d}$ are about a factor of 700 larger than the corresponding scalar mesons $B_{s,d}$. The obtained ratio of the branching fractions  $
Br({B_{s,d}^*\to \mu^+\mu^-})/{Br({B_{s,d}\to\mu^+\mu^-})}$ is about $\frac{0.3 \times {\rm eV}}{{\Gamma(B^*_{s,d} \to  B_{s,d} \gamma)}}$.  At the same time, the hadronic contribution
$B_{s,d} \to B^*_{s,d} \gamma \to \mu^+ \mu^-$ is estimated too. The relative increase of the amplitude of $B_{s,d}\to \mu^+\mu^-$ is about $(0.01\pm 0.006) \sqrt{\frac{{\Gamma(B^*_{s,d} \to  B_{s,d} \gamma)}}{{100~ {\rm eV}}}}$. If we choose $\Gamma(B^*_{s,d} \to  B_{s,d} \gamma)=2~$eV, the branching fractions of the vector mesons to lepton pair  are $(6.2 \pm 0.6) \times 10^{-10}$  and $(1.7 \pm 0.2) \times 10^{-11}$    for  $B^*_{s}$ and $B^*_{d}$ respectively. If we choose $\Gamma(B^*_{s,d} \to  B_{s,d} \gamma)=200~$eV, the updated branching fractions of the scalar mesons to muon pair are $(3.78 \pm 0.25)\times 10^{-9}$  and $(1.09 \pm 0.09)\times 10^{-10}$ for $B_{s}$ and $B_{d}$ respectively. Further studies on $B^*_{s,d}$ are usefully here, including dielectron decay, two-body decay with $J/\psi$, and so on.
}

\begin{document}
\maketitle
\flushbottom

%
%

\section{Introduction}

The leptonic decays of the $B_{s,d}$ mesons play an important role in the standard model (SM) and the new physics (NP) \cite{Bobeth:2013uxa,CMS:2014xfa}. 
They are highly suppressed in the SM since the flavor changing neutral current decays are
generated through W-box and Z-penguin diagrams. Furthermore, the
branching fractions of  leptonic decays of scalar meson  undergo an additional helicity suppression factor by
$m_\mu^2/M_{S}^2$, where $m_\mu$ and $M_{S}$ denote masses of the muon lepton and the scalar meson, respectively. The suppression factor  can be removed in some NP models, such as the  two-Higgs doublet  models    \cite{Cheng:2015yfu},
the minimal supersymmetric standard
model (MSSM)   \cite{Babu:1999hn}, the next minimal supersymmetric standard
model (NMSSM)   \cite{Li:2015dil}, the dark matter   \cite{Belanger:2015nma}, the universal extra dimensional model  \cite{Datta:2015aka}, lepton universality violation model  \cite{Glashow:2014iga}, the four generations fermion  \cite{Hou:2014nna},  and so on   \cite{Arbey:2012ax}.
The branching fractions of $B_{s,d} \to \mu^+ \mu^-$ measured by the CMS and LHCb Collaborations    \cite{CMS:2014xfa} and  predicted within the SM     \cite{Bobeth:2013uxa} with  NNLO QCD    \cite{Hermann:2013kca} and NLO EW    \cite{Bobeth:2013tba} corrections  included are collected in Table~\ref{tab:EXSM}.

Since the experimental branching fractions of $B_{s,d} \to \mu^+ \mu^-$ are measured from the dimuon distributions by the CMS and LHCb Collaborations  \cite{CMS:2014xfa}, the process $B^*_{s,d} \to \mu^+ \mu^-$~ will enhance the dimuon distributions for the mass splitting are about 45 MeV between $B_{s,d}$ and $B^*_{s,d}$. In another hand, the hadronic contribution
$B_{s,d} \to B^*_{s,d} \gamma \to \mu^+ \mu^-$  are missed in the  theoretical prediction  \cite{Bobeth:2013uxa}.
So we study $B^*_{s,d} \to \mu^+ \mu^-$~
 and its impact on $B_{s,d} \to \mu^+ \mu^-$ within SM here. The process $B_s \to B_s^* \gamma \to \mu^+ \mu^- \gamma$ was considered in Ref.~ \cite{Aditya:2012im}.
Recently, $B^*_{s,d} \to \mu^+ \mu^-$ are considered in Ref.    \cite{Grinstein:2015aua,Khodjamirian:2015dda}.
And hadronic contribution from charmonium  in $B \to K^{(*)} \ell^{+} \ell^{-}$ and $B \to X_s \gamma$
had been studied in Refs.  \cite{Falk:1993dh,Khodjamirian:2010vf}.

\begin{table}[t]
\caption{The branching fractions of  $B_{s,d} \to \mu^+ \mu^-$ measured by the CMS and LHCb Collaborations   \cite{CMS:2014xfa} and  predicted within the SM     \cite{Bobeth:2013uxa} with  NNLO QCD    \cite{Hermann:2013kca} and NLO EW   \cite{Bobeth:2013tba} corrections included\label{tab:EXSM}. }
\begin{center}
\begin{tabular}{cccc}
\hline
& EX  \cite{CMS:2014xfa}
& SM   \cite{Bobeth:2013uxa}
& Deviations
\\
\hline
$Br(B_{d}   \to   \mu^+   \mu^-)$  &  $(3.9 ^{+1.6}_{-1.4})  \times  10^{-10}$  &  $(1.06  \pm 0.09)   \times  10^{-10}$  &  $2.2\sigma$
\\
$Br(B_{s}   \to   \mu^+   \mu^-)$  &  $(2.8 ^{+0.7}_{-0.6})  \times  10^{-9}$  &  $(3.66  \pm 0.23)  \times   10^{-9}$  &  $1.2\sigma$
\\
$\frac{Br(B_{d} \to \mu^+ \mu^-)}
{Br(B_{s} \to \mu^+ \mu^-)}$  &  $0.14^{+0.08}_{-0.06}$  &  $0.0295^{+0.0028}_{-0.0025}$  &  $2.3\sigma$
\\
\hline
\end{tabular}
\end{center}
\end{table}

\section{The Decay of $B_s^* (B_d^*) \to \mu^+\mu^-$}

Within the SM, effective Lagrangian related  with $b  \bar{s} \to \mu^+ \mu^-$ 
is given in Ref.~ \cite{Grinstein:2004vb,Descotes-Genon:2013wba}
 \begin{eqnarray}\label{effham}
\mathcal{L} = \mathcal{N}\biggl[C_7^{eff}(\mu_f) \mathcal{O}^\gamma_7 + C_9 (\mu_f) \mathcal{O}^V_9+ C_{10} (\mu_f)\mathcal{O}^A_{10}\biggl],
\end{eqnarray}
where $\mathcal{N}=\frac{G_F}{\sqrt{2}} V_{tb} V^{*}_{ts}\frac{e^2}{4\pi^2}$, and the 
 operators $\mathcal{O}_{7,9,10}$ read as
 \begin{eqnarray}
\mathcal{O}^\gamma_7&=&- \frac{2im_b (p_\mu^\nu+p_{\bar\mu}^\nu )}{(p_\mu+p_{\bar\mu})^2} (\bar{s}\sigma_{\rho\nu}P_R b)(\bar \mu \gamma^{\rho} \mu),\\
\mathcal{O}^V_9&=& (\bar s\gamma_\rho P_Lb)(\bar \mu \gamma^{\rho} \mu),\\
\mathcal{O}^A_{10}&=& (\bar s\gamma_\rho P_Lb)(\bar\mu \gamma^{\rho}\gamma_5 \mu),
\end{eqnarray}
where $P_L=(1-\gamma_5)/2$~, $P_R=(1+\gamma_5)/2$. And the  Wilson coefficients are $C_{7, 9, 10}(\mu_f)=(-0.316, 4.403-0.47i, -4.493)$ at $\mu_f=m_b=4.5$~GeV~\cite{Khodjamirian:2015dda}. The superscript $\gamma$, $V$,  and $A$ denote the contributions from photon, vector current, and axial vector current, respectively.

The relations between the quark level operators 
and  the meson are described as
\begin{eqnarray}\label{Eq:decayconstant}
\langle 0|\bar s\gamma^\mu b|B_s^*(q,\varepsilon)\rangle     &=&   m_{B_s^*}\,f_{B_s^*}\,\varepsilon^\mu, \\
\langle 0|\bar s\sigma^{\mu\nu} b|B_s^*(q,\varepsilon)\rangle &=& -i\,f_{B_s^*}^T(q^\mu\varepsilon^\nu-\varepsilon^\mu q^\nu),
\\ \langle 0|\bar s\gamma^\mu \gamma_5 b|B_s(q)\rangle  &=&  -if_{B_s}\,q^\mu,
\end{eqnarray}
where the three decay constants $f_{B_s},~ f_{B_s^*}$ and $f_{B_s^*}^T$ depend on the renormalization scale. Their relations have been studied in Ref.~ \cite{Manohar:2000dt} in the heavy-quark limit. Ignoring the mass difference between $B_s$ and $B_s^*$ and the high order QCD corrections, we have
\begin{eqnarray}
f_{B_s^*}=f_{B_s^*}^T=\,f_{B_s}.
\end{eqnarray}

Then the amplitudes of $B_s^* (B_s) \to \mu^+\mu^-$~ are~ \cite{Aditya:2012im}
\begin{eqnarray}\label{eq:bsstarbamplitude}
\mathcal{M}({B_s^*\to \mu^+\mu^-})&=& f_{B_s^*}\frac{\mathcal{N}}{2} m_{B_s^*}\bar\mu \slash{\varepsilon}\Big [C_V^{eff}+C_{10}\gamma_5 \Big] \mu,   \nonumber \\
\mathcal{M}({B_s\to \mu^+\mu^-})&=&-if_{B_s}\mathcal{N}C_{10} m_\mu \bar\mu \gamma_5 \mu,
\end{eqnarray}
where
\begin{eqnarray}
C_V^{eff}=C_9+2\frac{m_b}{m_{B_s^*}} C_7^{eff}.
\end{eqnarray}
The helicity suppression factor $m_\mu^2/m_M^2$ in the decay width is removed in the vector meson decay.
Then we can get the decay widths of $B_s^* (B_s) \to \mu^+\mu^-$~
\begin{eqnarray}
\Gamma({B_s^*  \to   \mu^+\mu^-})&=&\frac{G_f^2\alpha_{em}^2}{96\pi^3}\left|V_{tb}V_{ts}^*\right|^2\Big(\left|C_{10}\right|^2
+\left|C_{V}^{eff}\right|^2)  m_{B_s^*}^3f_{B_s^*}^2\nonumber\\&&
\Big(1+{\cal{O}}(m_\mu^2/m_{B_s}^2)\Big) \nonumber \\
\Gamma({B_s  \to   \mu^+\mu^-})&=&\frac{G_f^2\alpha_{em}^2}{16\pi^3}\left|V_{tb}V_{ts}^*\right|^2\left|C_{10}\right|^2m_\mu^2m_{B_s}f_{B_s}^2\nonumber\\&&
\Big(1+{\cal{O}}(m_\mu^2/m_{B_s}^2)\Big)\\
\frac{
\Gamma({B_s^*  \to   \mu^+\mu^-})}
{\Gamma({B_s  \to   \mu^+\mu^-})}
&=&
\frac{\left(   \left|C_{10}\right|^2
+\left|C_{V}^{eff}\right|^2 \right)m_{B_s^*}^3f_{B_s^*}^2}{6\left|C_{10}\right|^2 m_\mu^2m_{B_s}f_{B_s}^2}
\Big(1  +  {\cal{O}}(m_\mu^2/m_{B_s}^2)\Big)\nonumber
\end{eqnarray}
The ratio of decay width is about 700 for $B_s^{(*)}$ and $B_d^{(*)}$ 
both.

\section{The impact of $B_s^* (B_d^*) \to \mu^+\mu^-$ on $B_s (B_d) \to B_s^* (B_d^*) \gamma \to \mu^+\mu^-$}

\begin{figure}[tbp]
\centering 
\includegraphics[width=.6\textwidth]{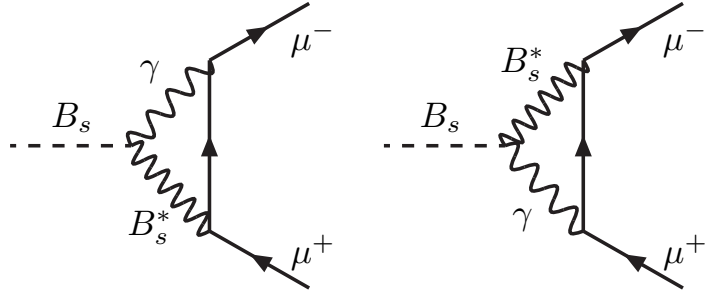}
\hfill
\caption{\label{fig:feymDiag1}  Feynman diagrams of $B_{s,d} \to  B^*_{s,d} \gamma^* \to \mu^+ \mu^-$.}
\end{figure}

Meanwhile, $B^*_{s,d}$ will impact on the leptonic decay of $B_{s,d}$~ through the loop contribution~ $B_{s,d} \to  B^*_{s,d} \gamma^* \to \mu^+ \mu^-$. The Feynman diagrams are given in Fig.~\ref{fig:feymDiag1}. The vertex of $B_{s,d} \to  B^*_{s,d} \gamma$ is given as the operator  \cite{Peskin:1995ev,Ebert:2002xz}
\begin{eqnarray}
\mathcal{M}_{B_s B_s^{*} \gamma}  &= &\sum_{q=s,b} <B_s^{*} \gamma| i e e_q \bar q (p_{\bar q}) \gamma_\mu q (p_q)| B_s> \nonumber \\
&= &\sum_{q=s,b}\varepsilon^{\mu}_\gamma p_\gamma^\nu<B_s^{*} | iee_q \bar q (p_{\bar q}) \frac{i \sigma_{\mu\nu}}{2m_q} q (p_q)| B_s> .\nonumber
\end{eqnarray}
We can simplify the matrix element $<B_s^{*} |  \bar q (p)  \sigma_{\mu\nu} q (p)| B_s> $  with the procedure in Refs.~ \cite{Beneke:2000wa,Cheung:2014cka}{\footnote{The decay constants defined of vector meson in Eq.~(\ref{Eq:decayconstant}) is different with Eq.(44) in Ref.~ \cite{Beneke:2000wa} with an additional $i$. }},
\begin{eqnarray}\label{eq:VPgammaVertex}
\mathcal{M}_{B_s B_s^{*} \gamma}  &=& \sum_{q=s,b}\frac{-e e_q}{2 m_q}\varepsilon^{\mu}_\gamma p_\gamma^\nu<B_s^{*} |  \bar q (p)  \sigma_{\mu\nu} q (p)| B_s> \nonumber
\\
& =&i
\epsilon_{\mu \nu \alpha \beta}
 \varepsilon^{\mu}_\gamma p^{\nu}_\gamma \varepsilon^{\alpha}_{B^*_s }    p^{\beta}_{B^*_s }\sum_{q=s,b}\left(\frac{e e_q}{ m_q}\right){\cal{I}}.
\end{eqnarray}
$\cal{I}$ is related with the wave functions of $B^*_s$ and $B_s$~ \cite{Beneke:2000wa}, and it is ${\cal{I}}=<B_s| j_0(p_\gamma r)| B_s^{*}>\sim 1$ in the non-relativistic limit  \cite{Lahde:1999ih}. We can rewrite Eq. (\ref{eq:VPgammaVertex}) with
\begin{eqnarray}
\mathcal{M}_{B_s B_s^{*} \gamma}  &=& i \frac{g_{B_s  B^*_s \gamma}}{m_{B_s^*}}  \epsilon_{\mu \nu \alpha \beta}
 \varepsilon^{\mu}_\gamma p^{\nu}_\gamma \varepsilon^{\alpha}_{B^*_s }    p^{\beta}_{B^*_s }
\end{eqnarray}
Where the  dimensionless vector-scalar-photon coupling constant $g_{B_s  B^*_s \gamma}$ is related with the magnetic moments of $b$ and $s$ quarks. And the phase factor $i$  is consistent with the amplitude of $\gamma^*\to VP$ in Ref. \cite{Ma:2004qf}.

There are ultraviolet (UV) logarithmical divergences  in the evaluation of loop integrals. Then we introduce a cut off regularization scheme for the UV divergence integral
\begin{eqnarray}
 &&\int \frac{d^4 q}{(2\pi)^4}\frac{1}{\left(q_i^2-m_i^2\right)\left(q_j^2-m_j^2\right)}
  \nonumber \\
  &\to&\int \frac{d^4 q}{(2\pi)^4}\left[\frac{1}{\left(q_i^2-m_i^2\right)\left(q_j^2-m_j^2\right)}\right.\nonumber \\
%
&&\left.-\frac{1}{\left(q_i^2-(m_i+\Lambda)^2\right)\left(q_j^2-(m_j+\Lambda)^2\right)}\right],
\end{eqnarray}
where $i,j= B_s^*,~\gamma$~or $\mu$, and $q_i$ corresponds the momentum of $i$ in loop. $\Lambda \ll M_W$ for the amplitude is UV finite when W boson are involved. Since the hadronic contribution will be suppressed when $\sqrt{q_j^2}-m_j\gg \Lambda_{QCD}$, $\Lambda$ is about several $\Lambda_{QCD}$. The cut off regularization scheme is similar  Pauli-Villars regularization scheme but acts on two propagators. 
The Pauli-Villars regularization scheme of the UV divergence integral is the same with the form factor ${\cal F}$ which is introduced in the $B_s  B^*_s \gamma$  vertex in Ref.~ \cite{Li:2007au}
\begin{eqnarray}
{\cal F} = \left(\frac {\Lambda^2 - m_{B^*_s}^2}
{\Lambda^2-q_{B^*_s}^2}\right),
\end{eqnarray}
for
\begin{eqnarray}
 \frac{1}{q_{B^*_s}^2-m_{B^*_s}^2}{\cal F}= \frac{1}{q_{B^*_s}^2-m_{B^*_s}^2}-\frac{1}{q_{B^*_s}^2-\Lambda^2}.
\end{eqnarray}
But here it acts on the UV divergence term only and the two propagators. Then the soft contribution will be  maintained in our calculation.

The amplitude from $B_s \to  B^*_s \gamma^* \to \mu^+ \mu^-$ can be written as
\begin{eqnarray}\label{eq:rn}
\mathcal{M}({B_s\to  B^*_s \gamma^* \to \mu^+\mu^-})
&=&i e \mathcal{N} g_{B_s  B^*_s \gamma}  R(\Lambda)
 C_V^{eff}f_{B_s^*}m_\mu  \bar\mu \gamma_5 \mu.
%
%
%
\end{eqnarray}
$m_\mu$ reappears in the amplitude of leptonic decay of scalar meson.
The factor $R(\Lambda)$ serves as a function of the high energy cut is shown in Fig.\ref{fig:RLambda}. More information about the factor $R(\Lambda)$ are given in the Appendix.

\begin{figure}[tbp]
\centering 
\includegraphics[width=.6\textwidth]{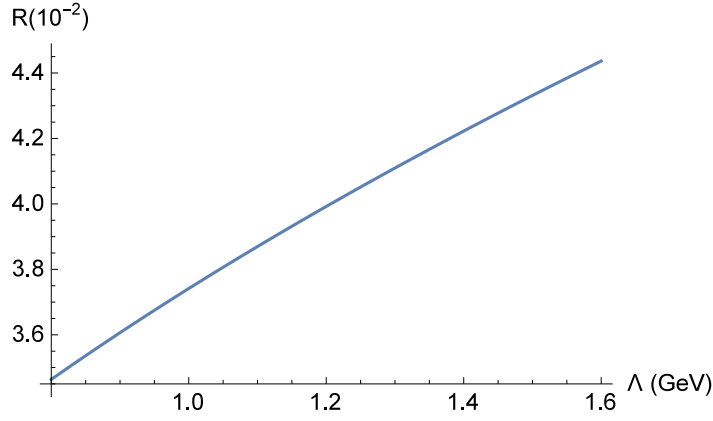}
\hfill
\caption{\label{fig:RLambda}  $R(\Lambda)$ of
$B_{s} \to  
\mu^+ \mu^-$ defined in
 Eq.(\ref{eq:rn}) as a function of the cut off energy.}
\end{figure}
Only the $C_{10}$ term is taken into account in the calculation of the NNLO QCD    \cite{Hermann:2013kca} and NLO EW   \cite{Bobeth:2013tba} corrections of $B_{s,d} \to \mu^+ \mu^-$ within the SM  \cite{Bobeth:2013uxa}. But the contribution from $B_s^*$   is
missed in the previous calculation, which is also related with $C_7$ and $C_9$ terms.

Compared with Eq.(\ref{eq:bsstarbamplitude}), the previous  amplitude is added a factor $F$,
\begin{eqnarray}\label{eq:relativeAMPFF}
F({B_s^*})&=& \frac{\mathcal{M}({B_s\to  B^*_s \gamma^* \to \mu^+\mu^-})}{\mathcal{M}({B_s\to \mu^+\mu^-})}\nonumber \\
&=&- \frac{C_V^{eff}f_{B_s^*}}{C_{10}f_{B_s}} {e g_{B_s  B^*_s \gamma}}  R(\Lambda)
\end{eqnarray}

We can estimate $g_{B_s  B^*_s \gamma}$ in several ways, including  the heavy-quark and chiral effective theories~ \cite{Cho:1992nt,Amundson:1992yp} with the radiative and pion transition widthes of $D^{*+}$, light cone QCD sum rules~ \cite{Zhu:1998ih,Zhu:1996qy}, and
the radiative
M1 decay widths of $B^*_s \to  B_s \gamma$ from potential model~ \cite{Lahde:1999ih,Goity:2000dk}.
The radiative M1 decay width of $B^*_s \to  B_s \gamma$ is
\begin{eqnarray}
g_{B_s  B^*_s \gamma}=-m_{ B^*_s}\left(\frac{12 \pi}{E_\gamma^3} \Gamma(B^*_s \to  B_s \gamma)\right)^{1/2}.
\end{eqnarray}
The predicted M1 widths are  $0.15 - 400~$eV and  $10 - 300~$eV   for  $B^*_s \to  B_s \gamma$ and   $B^*_d \to  B_d \gamma$, respectively \cite{Lahde:1999ih,Goity:2000dk,Ebert:2002xz,Cheung:2014cka,Cheung:2015rya}.

\section{Numerical Result}

The parameters in the numerical calculation are chosen as  \cite{Agashe:2014kda}
\begin{eqnarray}
\Lambda&=&1.2  ~{\rm GeV},\nonumber\\
m_b&=&4.2 ~{\rm GeV},\nonumber\\
\alpha_{em}&=&1/137.
\end{eqnarray}
The branch fraction of $B^*_{s,d}$  weak decay is much less than the M1 decay, $\Gamma_{tot}(B^*_{s,d})\approx \Gamma(B^*_{s,d} \to  B_{s,d} \gamma)$.
We can get the ratio
%
\begin{eqnarray}
\frac{
Br({B_s^*\to \mu^+\mu^-})}
{Br({B_s\to\mu^+\mu^-})}&=&(0.34 \pm 0.03) \times \frac{\rm eV}{\Gamma(B^*_{s} \to  B_{s} \gamma)},
\nonumber\\
\frac{
Br({B_d^*\to \mu^+\mu^-})}
{Br({B_d\to\mu^+\mu^-})}&=&(0.33 \pm 0.03)\times  \frac{\rm eV}{\Gamma(B^*_{d} \to  B_{d} \gamma)}.
\end{eqnarray}
The main uncertainty resulting from the value of $f_{B_{s,d}}^*$.
The distribution of the dimuon invariant mass measured by CMS and LHCb Collaborations  in Ref.~ \cite{CMS:2014xfa} should include the contributions of $B^*_{s,d}\to \mu^+ \mu^-$.
If $\Gamma(B^*_{s} \to  B_{s} \gamma)=1~$eV~ \cite{Lahde:1999ih}, we can get
$
\frac{
Br({B_s^*\to \ell^+\ell^-})}
{Br({B_s\to\mu^+\mu^-})}=0.34
$ for  $\ell=e,~\mu$. Then 
$B_s^*\to e^+e^-$ may be searched by CMS and LHCb experiments with larger data samples.

In another hand, if $\Gamma(B^*_{s,d} \to  B_{s,d} \gamma)\sim 100$ eV, we can find that the amplitude of
$B_{s,d}\to \mu^+\mu^-$ will be  modified by the contributions of $B_{s,d}^*$  with a  factor
\begin{eqnarray}\label{eq:relativeAMPnumber}
F({B_{s,d}^*})&=&(0.011\pm 0.006) \sqrt{\frac{\Gamma(B^*_{s,d} \to  B_{s,d} \gamma)}{100 {\rm eV}}} ,
\end{eqnarray}
The main uncertainty resulting from the value of $\Lambda$.
Then the new predictions of $\Gamma(B_{s,d} \to \mu^+\mu^-)$ are
\begin{eqnarray}
Br(B_{s}  \to   \mu^+ \mu^-)&=&(36.6
\pm    2.3 )   \times    \left(1  +   ( 0.023 \pm 0.012) \sqrt{\frac{\Gamma_{tot}(B^*_{s} )}{100 {\rm eV}}}\right)  \times
10^{-10}\nonumber,\\
Br(B_{d}  \to   \mu^+ \mu^-)&=&(10.6    \pm     0.9)  \times
\left(1  +   ( 0.023 \pm 0.012)\sqrt{\frac{\Gamma_{tot}(B^*_{d} )}{100 {\rm eV}}}\right) \times   10^{-11}.\nonumber
\end{eqnarray}
If $\Gamma(B^*_{s,d} \to  B_{s,d} \gamma)=200$~eV, this factor will increase the decay width $\Gamma(B_{s,d} \to \mu^+\mu^-)$ by a factor $(3.3 \pm 1.7)\%$, which is about a factor of 10 larger than the neglect NLO EW correction factor $0.3\%$ at the decay width in Ref. \cite{Bobeth:2013uxa}. And the corresponded  $g_{B_{s,d}  B^*_{s,d} \gamma}=-1.5$, about a factor of 15 larger than the $e_q e=-1/3\sqrt{4\pi \alpha_{em}}=-0.10$.
The $\Gamma(B^*_{s,d} \to  B_{s,d} \gamma)$ may be measured through two-body decay $B^*_{s,d}\to J/\psi +M$ by CMS and LHCb. 

\section{Summary}

In summary, $B^*_{s,d}\to \mu^+ \mu^-$ in the dimuon distributions and  the hadronic contribution
$B_{s,d} \to B^*_{s,d} \gamma \to \mu^+ \mu^-$  are  studied here. The $ \mu^+ \mu^-$ decay widths of vector mesons $B^*_{s,d}$ are about a factor of 700 larger than the corresponding scalar mesons $B_{s,d}$. The obtained ratio of the branching fractions  $
Br({B_{s,d}^*\to \mu^+\mu^-})/{Br({B_{s,d}\to\mu^+\mu^-})}$ is about $0.3 \times {\rm eV}/{\Gamma(B^*_{s,d} \to  B_{s,d} \gamma)}$.  At the same time, the hadronic contribution
$B_{s,d} \to B^*_{s,d} \gamma \to \mu^+ \mu^-$ is calculated too. The amplitude of $B_{s,d}\to \mu^+\mu^-$ is enhanced by a factor of $0.01 \sqrt{{\Gamma(B^*_{s,d} \to  B_{s,d} \gamma)}/{100~ {\rm eV}}}$. If we choose $\Gamma(B^*_{s,d} \to  B_{s,d} \gamma)=2~$eV, the branching fractions of the vector mesons to lepton pair  are $5.3 \times 10^{-10}$  and  $1.6 \times 10^{-11}$ for  $B^*_{s}$ and $B^*_{d}$ respectively. If we choose $\Gamma(B^*_{s,d} \to  B_{s,d} \gamma)=200~$eV, the updated branching fractions of the scalar mesons to muon pair are $(3.78 \pm 0.25)\times 10^{-9}$  and $(1.09 \pm 0.09)\times 10^{-10}$ for $B_{s}$ and $B_{d}$ respectively. Further studies  on $B^*_{s,d}$ are usefully here, including dielectron decay, two-body decay with $J/\psi$, and so on.

\acknowledgments{We would like to thank K.T. Chao, Y.Q. Ma, C. Meng, Q. Zhao, and S.L. Zhu for useful discussions.
This work is supported by the National Natural Science
Foundation of China (Grants No. 11375021, 11575017), the New Century Excellent Talents in University (NCET) under grant
NCET-13-0030,  the Major State Basic Research Development Program of China (No. 2015CB856701),  and
the Fundamental Research Funds for the Central Universities.
}

\section{Appendix: $R(\Lambda)$}

$R(\Lambda)$ of
$B_{s} \to  B^*_{s} \gamma^* \to
\mu^+ \mu^-$ defined in
 Eq.(\ref{eq:rn}) is given as:

\begin{eqnarray}
R(\Lambda)&=&\frac{1}{32 \pi ^2 {m_{B_s}}^2 m_\mu^2}\left\{ 3 {m_{B_s}}^2
   m_\mu^2 -2m_\mu^2 \left({m_{B_s}}^2-
   {m_{B^*_s}}^2\right)^2 {C_0}\left({m_{B_s}}^2,m_\mu^2,m_\mu^
   2,{m_{B^*_s}}^2,0,m_\mu^2\right) \right.\nonumber \\&&
   \left.
   +{m_{B_s}}^2(m_\mu^2-{m_{B^*_s}}^2)\left({B_0}\left(0,m_\mu^2,{m_{B^*_s}}^2\right)
   -{B_0}\left(0,({\Lambda}+m_\mu)^2,(
   {\Lambda}+{m_{B^*_s}})^2\right)\right)\right.\nonumber \\
   &&
  \left. +\left(m_{B_s}^2 \left(2m_\mu^2 +{m_{B^*_s}}^2\right)+2 m_\mu^2{m_{B^*_s}}^2\right) \left({B_0}\left(m_\mu^2,m_\mu^2,{m_{B^*_s}}^2\right)-{B_0}\left(m_\mu^2,
  (\Lambda+m_\mu)^2,({\Lambda}+{m_{B^*_s}})^2\right)\right)\right.\nonumber \\
  &&
   \left.
   +m_\mu^2 \left(3 {m_{B_s}}^2-2{m_{B^*_s}}^2\right) \left({B_0}\left(m_\mu^2,0,m_\mu^2\right)-
   {B_0}\left(m_\mu^2,{\Lambda}^2,(\Lambda+m_\mu)^2\right)\right) 
 \right\}
\end{eqnarray}
The scalar functions ${B_0}$ and ${C_0}$ are given in Ref.~\cite{'tHooft:1978xw,vanOldenborgh:1989wn,vanOldenborgh:1990yc}. As a numerical fit between $0.5-2$~GeV, we can get
\begin{eqnarray}
R(\Lambda)=0.022+0.062\times ln(\frac{\Lambda+m_{B_s}}{m_{B_s}}).
\end{eqnarray}


\providecommand{\href}[2]{#2}\begingroup\raggedright\endgroup

\end{document}